\documentstyle[preprint,aps,floats,epsf]{revtex}

\begin{document}

\tightenlines 

\newcommand{\notE}{\ \hbox{{$E$}\kern-.60em\hbox{/}}}
\newcommand{\notp}{\ \hbox{{$p$}\kern-.43em\hbox{/}}}


\includegraphics{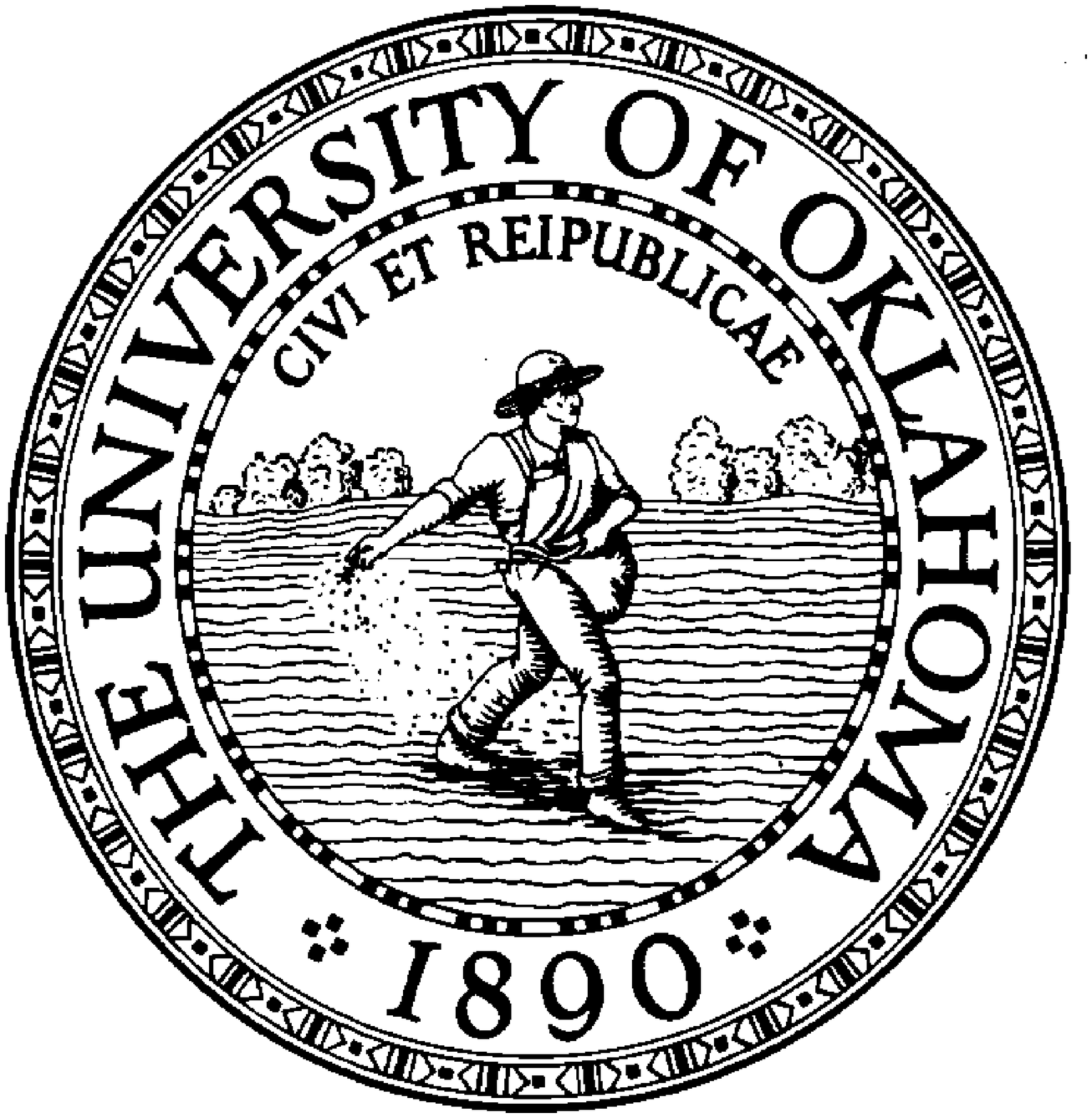}

\preprint{\font\fortssbx=cmssbx10 scaled \magstep2
\hbox to \hsize{
\hskip1.2in 
\hbox{\fortssbx The University of Oklahoma}
\hskip0.8in $\vcenter{
                      \hbox{\bf BNL-HET-02/15}
                      \hbox{\bf OKHEP-02-02}
                      \hbox{\bf UTHEP-02-06}
                      \hbox{\bf hep-ph/0208063}
                      \hbox{August 2002}}$ }
}
 
\title{\vspace{.5in}
Searching for the Higgs Bosons of Minimal Supersymmetry \\ 
with Muon Pairs and Bottom Quarks}
 
\author{Sally Dawson$^a$, Duane Dicus$^b$ and Chung Kao$^c$}

\address{
$^a$Department of Physics, Brookhaven National Laboratory, 
Upton, NY 11973, USA \\
$^b$Center for Particle Physics, University of Texas, 
Austin, TX 78712, USA \\
$^c$Department of Physics and Astronomy, University of Oklahoma, 
Norman, OK 73019, USA}

\maketitle

\bigskip

\begin{abstract}

The prospects for the discovery of neutral Higgs bosons 
($\phi^0 = H^0, h^0, A^0$) produced with 
bottom quarks via Higgs decays into muon pairs 
($pp \to b\bar{b}\phi^0 \to b\bar{b}\mu\bar{\mu} +X$) 
at the CERN LHC 
are investigated in the minimal supersymmetric model.  
The complete physics background from the production of 
$b\bar{b}\mu\bar{\mu}$, $b\bar{b}W^+W^-$ (including $t\bar{t}$) 
and $jj\mu\bar{\mu}, j = g, u, d, s, c$ in the Standard Model 
is calculated with realistic acceptance cuts. 
This discovery mode has a simple production mechanism from 
$gg \to b\bar{b}\phi^0$ with its cross section 
proportional to $1/\cos^2\beta$ and could provide an opportunity 
to measure $\tan\beta$ and the $b\bar{b}\phi^0$ couplings.
In addition, we compare the associated discovery mode above with 
the inclusive discovery channel $pp \to \phi^0 \to \mu\bar{\mu} +X$.
Promising results are found for the CP-odd pseudoscalar ($A^0$) 
and the heavier CP-even scalar ($H^0$) Higgs bosons 
for $\tan\beta \equiv v_2/v_1 \agt 14$ and $m_A,m_H \alt$ 325 GeV.

\end{abstract}

%

\newpage

\section{Introduction}

In the Standard Model (SM), 
the masses of gauge bosons and fermions are generated by one scalar field 
doublet. After spontaneous symmetry breaking, 
a neutral CP-even Higgs boson ($h^0_{SM}$) remains as a physical particle.

A supersymmetry (SUSY) between fermions and bosons provides 
a natural explanation of the Higgs mechanism 
for electroweak symmetry breaking (EWSB) from radiative corrections 
in the framework of a grand unified theory 
and preserves the elementary nature of Higgs bosons. 
The Higgs sector of a supersymmetric theory must contain 
at least two $SU(2)$ doublets \cite{Guide} for anomaly cancellation. 
In the minimal supersymmetric standard model (MSSM) \cite{MSSM}, 
the Higgs sector has two doublets $\phi_1$ and $\phi_2$ 
that couple to the $t_3 = -1/2$ and $t_3 = +1/2$ fermions, respectively.  
After spontaneous symmetry breaking, there remain five physical Higgs bosons:
a pair of singly charged Higgs bosons $H^{\pm}$,
two neutral CP-even scalars $H^0$ (heavier) and $h^0$ (lighter),
and a neutral CP-odd pseudoscalar $A^0$.
The Higgs potential is constrained by supersymmetry 
such that all tree-level Higgs boson masses and couplings 
are determined by just two independent parameters,  
commonly chosen to be the mass of the CP-odd pseudoscalar ($m_A$) 
and the ratio of vacuum expectation values (VEVs) of the neutral Higgs fields 
($\tan\beta \equiv v_2/v_1$). 

In the Model II of two Higgs doublet models \cite{Model2} and in the MSSM, 
the couplings and production cross sections of 
$\phi^0 b\bar{b}, \phi^0 = H^0, h^0, A^0$ 
are enhanced by factors of $1/\cos\beta$.
For large $\tan\beta$, 
the $\tau\bar{\tau}$ decay mode \cite{Kunszt,Richter-Was}
is a promising discovery channel for the $A^0$ and the $H^0$ 
at the CERN Large Hadron Collider (LHC).
It has also been suggested that neutral Higgs bosons might be observable via
their $b\bar{b}$ decays \cite{Dandi,DGV} in a large region
of the ($m_A,\tan\beta$) plane,
provided that sufficient $b$-tagging capability can be achieved.
However, simulations for the ATLAS detector concluded that 
detection of the $b\bar{b}$ channel would be difficult \cite{hbb}.

The LHC discovery potential of the muon pair channel for neutral Higgs bosons 
in the minimal supersymmetric model 
was demonstrated by Kao and Stepanov \cite{Nikita,CMS},  
and was later confirmed by the ATLAS collaboration \cite{Richter-Was,ATLAS}. 
In the minimal supergravity unified model \cite{mSUGRA}, 
the significance of $pp \to \phi^0 \to \mu\bar{\mu} +X$ is 
greatly improved at large $\tan\beta$ \cite{Vernon} because  
$m_A$ and $m_H$ become small from the evolution of 
renormalization group equations with large $b\bar{b}\phi^0$ couplings.
On the other hand, it is very challenging to search for 
muon pairs from Higgs decays in the Standard Model 
\cite{Tilman,Han}.

Although the $\mu\bar{\mu}$ channel has a small branching fraction, 
this is compensated by an improved mass resolution for the muon pairs.
For large $\tan\beta$, the muon pair discovery mode might 
be the only channel at the LHC that allows precise reconstruction
of the $A^0$ and the $H^0$ masses. 

In this article, we present the prospects for discovering 
the MSSM neutral Higgs bosons ($\phi^0 = H^0, h^0, A^0$) 
produced with bottom quarks via Higgs decays into muon pairs 
($pp \to b\bar{b}\phi^0 \to b\bar{b} \mu\bar{\mu} +X$) at the LHC.
We calculate the Higgs signal and complete SM background 
of $b\bar{b} \mu\bar{\mu}$ with realistic cuts 
and compare the discovery potential 
with that of the inclusive final state of 
$pp \to \phi^0 \to \mu\bar{\mu} +X$. 
The production cross sections and branching fractions of Higgs bosons 
are discussed in Section II.
The dominant physics backgrounds from production of $b\bar{b}\mu\bar{\mu}$, 
and $b\bar{b} W^+ W^-$ are presented in Section III.
The observability of the dimuon signal is discussed in Section IV 
for conservative assumptions about the detector mass resolution. 
Conclusions are drawn in Section V.

\section{The production cross sections and branching fractions}

We calculate the cross section at the LHC for 
$pp \to b\bar{b} \phi^0 +X$ ($\phi^0 = H^0, h^0, A^0$) from two subprocesses 
$gg \to b\bar{b} \phi^0$ \cite{Duane} and $q\bar{q} \to b\bar{b} \phi^0$ 
with the parton distribution functions of CTEQ5L \cite{CTEQ5}. 
It turns out that the contribution from quark-antiquark annihilation 
($q\bar{q} \to b\bar{b} \phi^0$) is negligible.


In the MSSM, gluon fusion ($gg \to \phi^0$) is the major source
of neutral Higgs bosons for $\tan\beta$ less than about 4.
If $\tan\beta$ is larger than 7,
neutral Higgs bosons in the MSSM are dominantly produced
from ($gg \to b\bar{b} \phi^0$) \cite{Duane}. 
Since the Yukawa couplings of $\phi^0 b\bar{b}$ are enhanced by $1/\cos\beta$,
the production rate of neutral Higgs bosons is usually
enhanced at large $\tan\beta$.
For $m_A$ larger than about 150 GeV, the couplings of the lighter scalar $h^0$
to gauge bosons and fermions become similar to those of the SM Higgs boson. 
In this case, gluon fusion is the major source of the $h^0$ 
even if $\tan\beta$ is large.


In general, the QCD radiative corrections to the subprocess $gg \to \phi^0$ 
are substantial \cite{QCD1,QCD2,QCD3,QCD4}.
However, the next to the leading order (NLO) cross section of 
$pp \to \phi^0 b\bar{b} +X$ via $gg \to \phi^0 b\bar{b}$ 
with QCD radiative corrections has been found to be only slightly reduced 
from the leading order (LO) contribution 
evaluated with the bottom-quark pole mass $M_b$ \cite{Plumper}.
It is a good approximation to evaluate 
the NLO cross section of $pp \to \phi^0 b\bar{b} +X$ 
with the leading order (LO) contribution 
multiplied by a K factor\footnote{
The K factor is defined as $K = \sigma_{\rm NLO}/\sigma_{\rm LO}$, 
where $\sigma_{LO}$ is evaluated with LO parton distribution functions and
1-loop evolution of the strong coupling ($\alpha_s$) and
$\sigma_{NLO}$ is evaluated with NLO parton distribution functions and
2-loop evolution of $\alpha_s$. This is consistent with the notation
of Refs. \cite{QCD2} and \cite{Plumper}.} 
of 0.8 \cite{Plumper}  and using the pole mass $M_b = 4.7$ GeV. 

The results of Ref. \cite{QCD2} are applied to 
compute the NLO rates for $pp \rightarrow \phi^0$.
We take a K factor of 1.5 for the contribution from $gg \to H^0,h^0$ 
for $\tan\beta < 6$ and a K factor of 1.3 for $\tan\beta \ge 6$.
The contribution from $gg \to A^0$ is calculated with the 
LO contribution multiplied by a K factor of 1.7
for $\tan\beta < 6$ and a K factor of 1.3 for $\tan\beta \ge 6$.


The cross section for $pp \to b\bar{b} \phi^0 \to b\bar{b} \mu\bar{\mu} +X$ 
is evaluated with the Higgs production cross section 
$\sigma(pp \to b\bar{b} \phi^0 +X)$ 
multiplied by the branching fraction of the Higgs decay into muon pairs
$B(\phi^0 \to \mu\bar{\mu})$.
With QCD radiative corrections to $\phi^0 \to b\bar{b}$ \cite{Manuel}, 
the branching fraction of $\phi^0 \to \mu\bar{\mu}$ is about 
$m_\mu^2/3 m_b^2(m_\phi)$
when the $b\bar{b}$ mode dominates Higgs decays, 
where 3 is a color factor of the quarks 
and $m_b(m_\phi)$ is the running mass at the scale $m_\phi$.  
This results in the branching fraction for $A^0 \to \mu\bar{\mu}$ 
of approximately $3 \times 10^{-4}$ for $m_A = 100$ GeV.

\section{The Physics Background}
 
The dominant physics backgrounds to the final state of $b\bar{b}\mu\bar{\mu}$ 
come from $gg \to b\bar{b}\mu\bar{\mu}$ and $q\bar{q} \to b\bar{b}\mu\bar{\mu}$
as well as $gg \to b\bar{b}W^+W^-$ and $q\bar{q} \to b\bar{b}W^+W^-$
followed by the decays of $W^\pm \to \mu^\pm \nu_\mu$.
The Feynman diagrams for $gg \to b\bar{b}\mu\bar{\mu}$ are 
shown in Fig. 1.
We have also considered backgrounds from 
$pp \to gb \mu\bar{\mu} +X$, $pp \to g\bar{b} \mu\bar{\mu} +X$, 
$pp \to gq \mu\bar{\mu} +X$, $pp \to g\bar{q} \mu\bar{\mu} +X$, and 
$pp \to jj \mu\bar{\mu} +X$, 
where $q = u, d, s, c$ and $j = g, q$ or $\bar{q}$.

Our acceptance cuts and efficiencies of $b$-tagging and mistagging  
are the same as those of the ATLAS collaboration \cite{ATLAS}.
In each event, two isolated muons are required to have
$p_T(\mu) > 20$ GeV and $|\eta(\mu)| < 2.5$. 
For an integrated luminosity ($L$) of 30 fb$^{-1}$, 
we require $p_T(b,j) > 15$ GeV and $|\eta(b,j)| < 2.5$. 
The $b$-tagging efficiency ($\epsilon_b$) is taken to be $60\%$, 
the probability that a $c$-jet is mistagged as a $b$-jet ($\epsilon_c$)
is $10\%$ and 
the probability that any other jet is mistagged as a $b$-jet ($\epsilon_j$)
is taken to be $1\%$.
For $m_\phi < 100$ GeV, we change the requirement to $p_T(\mu) >$ 10 GeV 
for muons in both the Higgs signal and the background.

For a higher integrated luminosity of 300 fb$^{-1}$, 
we require the same acceptance cuts as those for $L =$ 30 fb$^{-1}$ 
except $p_T(b,j) > 30$ GeV and $\epsilon_b =50\%$.
In addition, we require that 
the missing transverse energy ($\notE_T$) in each event 
should be less than 20 GeV for $L = 30$ fb$^{-1}$ 
and less than 40 GeV for $L = 300$ fb$^{-1}$ 
to reduce the background from 
$pp \to b\bar{b}W^+W^- +X$ which receives its major contribution 
from both real and virtual top quarks
$pp \to t^* \bar{t}^* +X$.

We have employed the programs MADGRAPH \cite{Madgraph}
and HELAS \cite{Helas} to evaluate
the background cross sections of 
$pp \to b\bar{b}\mu\bar{\mu} +X, jj\mu\bar{\mu} +X$ and $bbW^+W^- +X$.
In Figure 2, we present distributions for the transverse momenta ($p_T$) 
of muons and bottom quarks 
for $pp \to b\bar{b} A^0 \to b\bar{b} \mu\bar{\mu} +X$ 
and  for the SM background of $pp \to b\bar{b} \mu\bar{\mu} +X$ via
$gg \to b\bar{b} \mu\bar{\mu}$ and 
$q\bar{q} \to b\bar{b} \mu\bar{\mu}$.
It is clear that the muon $p_T$ cut is effective in removing 
most of the SM background, while most muons from the Higgs decays 
pass the $p_T$ cut.
Most $b$ quarks in both the signal and the background do not have large $p_T$, 
therefore, only approximately 10$\%$ of the $b\bar{b}\mu\bar{\mu}$ 
events survive the requirement of $p_T(b,j) > 15$ GeV for $m_\phi = 100$ GeV.

We have also studied the inclusive discovery channel of 
$pp \to \phi^0 \to \mu\bar{\mu} +X$.
The dominant physics background to the inclusive final state of $\mu\bar{\mu}$ 
comes from the Drell-Yan process $q\bar{q} \to Z,\gamma \to \mu\bar{\mu}$ 
\cite{Nikita}. 
We make the same cuts discussed above and veto events 
with any jet passing the cuts on $p_T$ and $\eta$ 
to effectively reduce additional backgrounds from 
$pp \to t\bar{t} \to bb W^+ W^- +X$ and $pp \to jj \mu\bar{\mu} +X$ 
where $j = g, u, d, s, c$ or $b$.
 
\section{The Discovery Potential at the LHC}

We show the invariant mass distribution of muon pairs in Fig. 3 
for $pp \to b\bar{b} A^0 +X \to b\bar{b} \mu\bar{\mu} +X$ 
with $\tan\beta = 10$ and 50, 
the SM processes of $pp \to b\bar{b} \mu\bar{\mu} +X$, and
contributions from $pp \to b\bar{b} W^+ W^- +X$ 
with cuts and tagging efficiencies for $L =$ 30 fb$^{-1}$.
Also shown are the muon pair invariant mass distributions 
for the inclusive final state of $pp \to A^0 \to \mu\bar{\mu} +X$
and its dominant background from the Drell-Yan process
$pp \to Z,\gamma \to \mu\bar{\mu} +X$.
There are several interesting aspects to note from this figure: 
(a) including two $b$ quarks in the final state greatly improves the 
signal to background ratio with realistic efficiencies of b tagging 
and jet mistagging at the ATLAS and the CMS detectors,
(b) the SM subprocess of $gg \to b\bar{b} \mu\bar{\mu}$ and 
$q\bar{q} \to b\bar{b} \mu\bar{\mu}$ make the major contribution 
to the physics background for $M_{\mu\bar{\mu}} \alt 100$ GeV,
but $gg \to b\bar{b}W^+W^-$ and $q\bar{q} \to b\bar{b}W^+W^-$ 
become the dominant background for higher muon pair invariant mass. 

To study the discovery potential of 
$pp \to b\bar{b} \phi^0 +X \to b\bar{b} \mu\bar{\mu} +X$ at the LHC, 
we calculate the background from the SM processes of 
$pp \to b\bar{b} \mu\bar{\mu} +X$
in the mass window of
$m_\phi \pm \Delta M_{\mu\bar{\mu}}$ where 
$\Delta M_{\mu\bar{\mu}} \equiv 
1.64 [ (\Gamma_\phi/2.36)^2 +\sigma_m^2 ]^{1/2}$ \cite{ATLAS},
$\Gamma_\phi$ is the total width of the Higgs boson,  
and $\sigma_m$ is the muon mass resolution.
We take $\sigma_m$ to be $2\%$ of the Higgs boson mass \cite{ATLAS}.
The CMS mass resolution will be better than $2\%$ of $m_\phi$ for 
$m_\phi \alt$ 500 GeV \cite{Nikita,CMS}. 
Therefore, the observability for the muon pair discovery channel 
at the CMS detector will be better than what is shown in this article.

To study the observability of the inclusive muon discovery mode,
we consider the background from the Drell-Yan process,
$q\bar{q} \to Z,\gamma \to \mu\bar{\mu}$, which is the dominant background 
\cite{Nikita,CMS,ATLAS}. 
We require (i) $p_T(\mu) > 20$ GeV and 
(ii) $|\eta(\mu)| < 2.5$ for both the signal and background.

We define the signal to be observable 
if the $N\sigma$ lower limit on the signal plus background is larger than 
the corresponding upper limit on the background \cite{HGG,Brown}, namely,
\begin{eqnarray}
L (\sigma_s+\sigma_b) - N\sqrt{ L(\sigma_s+\sigma_b) } > 
L \sigma_b +N \sqrt{ L\sigma_b }
\end{eqnarray}
which corresponds to
\begin{eqnarray}
\sigma_s > \frac{N^2}{L} \left[ 1+2\sqrt{L\sigma_b}/N \right]
\end{eqnarray}
Here $L$ is the integrated luminosity, 
$\sigma_s$ is the cross section of the Higgs signal, 
and $\sigma_b$ is the background cross section 
within a bin of width $\pm\Delta M_{\mu\bar{\mu}}$ centered at $m_\phi$. 
In this convention, $N = 2.5$  corresponds to a 5$\sigma$ signal.
We take the integrated luminosity $L$ to be 30 fb$^{-1}$ 
and 300 fb$^{-1}$ \cite{ATLAS}. 

For $\tan\beta \agt 10$, 
$m_A$ and $m_H$ are almost degenerate when $m_A \agt$ 125 GeV, 
while $m_A$ and $m_h$ are very close 
to each other for $m_A \alt$ 125 GeV in the MSSM \cite{Nikita,CMS}.
Therefore, when computing the discovery reach, we add the cross sections 
of the $A^0$ and the $h^0$ for $m_A < 125$ GeV 
and those of the $A^0$ and the $H^0$ for $m_A \ge 125$ GeV.

The 5$\sigma$ discovery contours for the MSSM Higgs bosons at 
$\sqrt{s} =$ 14 TeV 
for an integrated luminosity of $L = 30 \;\; {\rm fb}^{-1}$ and 
$L = 300 \;\; {\rm fb}^{-1}$ are shown in Fig. 4.
We have chosen 
$M_{\rm SUSY} = m_{\tilde{q}} = m_{\tilde{g}} = m_{\tilde{\ell}} 
= \mu = 1$ TeV.
The discovery region is the part of the parameter space 
above the $5\sigma$ contour.

We have presented results for $M_{\rm SUSY} = 1$ TeV.
If $M_{\rm SUSY}$ is smaller, 
the discovery region of $A^0,H^0 \to \mu\bar{\mu}$ 
will be slightly reduced for $m_A \agt 250$ GeV,
because the the Higgs bosons can decay into SUSY particles \cite{HZ2Z2} 
and the branching fraction of $\phi^0 \to \mu\bar{\mu}$ is suppressed.
For $m_A \alt 125$ GeV,
the discovery region of $H^0 \to \mu\bar{\mu}$ 
is slightly enlarged for a smaller $M_{\rm SUSY}$, 
but the observable region of $h^0 \to \mu\bar{\mu}$ is slightly reduced 
because the lighter top squarks make the $H^0$ and the $h^0$ lighter; 
also the $H^0 b\bar{b}$ coupling is enhanced 
while the $h^0 b\bar{b}$ coupling is reduced \cite{Nikita,Vernon}.

\section{Conclusions}
 
The muon pair decay mode is a promising channel for the discovery of 
the neutral Higgs bosons in the minimal supersymmetric model at the LHC. 
The $A^0$ and the $H^0$ should be observable in a large region 
of parameter space with $\tan\beta > 10$.
The $h^0$ should be observable in a region with $m_A \alt 125$ GeV
and $\tan\beta > 5$.

In particular, Fig. 4 shows that the associated final state of 
$b\bar{b}\phi^0 \to b\bar{b}\mu\bar{\mu}$ 
could discover the $A^0$ and the $H^0$ at the LHC 
with an integrated luminosity of 30 fb$^{-1}$ if $m_A \alt 300$ GeV.
At a higher luminosity of 300 fb$^{-1}$, the discovery region in $m_A$ 
is not expanded much; the harder $p_T$ cut on $b$ quarks reduces 
the Higgs production cross section, and the higher missing transverse energy
slightly increases the background from $b\bar{b}W^+W^-$.

The inclusive final state of $\phi^0 \to \mu\bar{\mu}$ 
could also allow the discovery of the $A^0$ and the $H^0$ at the LHC 
with an integrated luminosity of 30 fb$^{-1}$ if $m_A \alt 450$ GeV.  
At a higher luminosity of 300 fb$^{-1}$, the statistical significance 
of the Higgs signal is improved by a factor of $\sqrt{10}$ and 
the discovery region in $m_A$ is significantly extended to $m_A \alt 650$ GeV 
for $\tan\beta = 50$.

The discovery of both 
$\phi^0 \to \tau\bar{\tau}$ and $\phi^0 \to \mu\bar{\mu}$ 
will allow us to understand the Higgs Yukawa couplings with the leptons.
The discovery of the associated final state of 
$b\bar{b}\phi^0 \to b\bar{b}\mu\bar{\mu}$ 
will provide information about the Yukawa couplings of 
$b\bar{b}\phi^0$ and an opportunity to measure $\tan\beta$. 
Although the discovery region of the $\mu\bar{\mu}$ mode 
is smaller than the $\tau\bar{\tau}$ channel, 
the $\mu\bar{\mu}$ channel allows a precise reconstruction
of the Higgs boson masses. 
Furthermore, if $m_H-m_A$ is smaller than 
the tau mass resolution, it would be difficult to separate them 
in the tau pair discovery mode.  
The muon discovery channel at the LHC might be employed 
to distinguish the $H^0$ and the $A^0$ in multi-Higgs-doublet models 
for $m_H-m_A$ slightly larger than the muon mass resolution ($\sigma_m$) 
of the CMS and the ATLAS detectors. 

\section*{Acknowledgments}

We are grateful to Howie Haber, Michael Spira and Scott Willenbrock 
for beneficial discussions. 
This research was supported in part by the U.S. Department of Energy
under Grants 
No.~DE-AC02-76CH00016, 
No.~DE-FG03-98ER41066, and 
No.~DE-FG03-93ER40757.
 
%



\begin{figure}
\centering\leavevmode
\epsfxsize=6in\epsffile{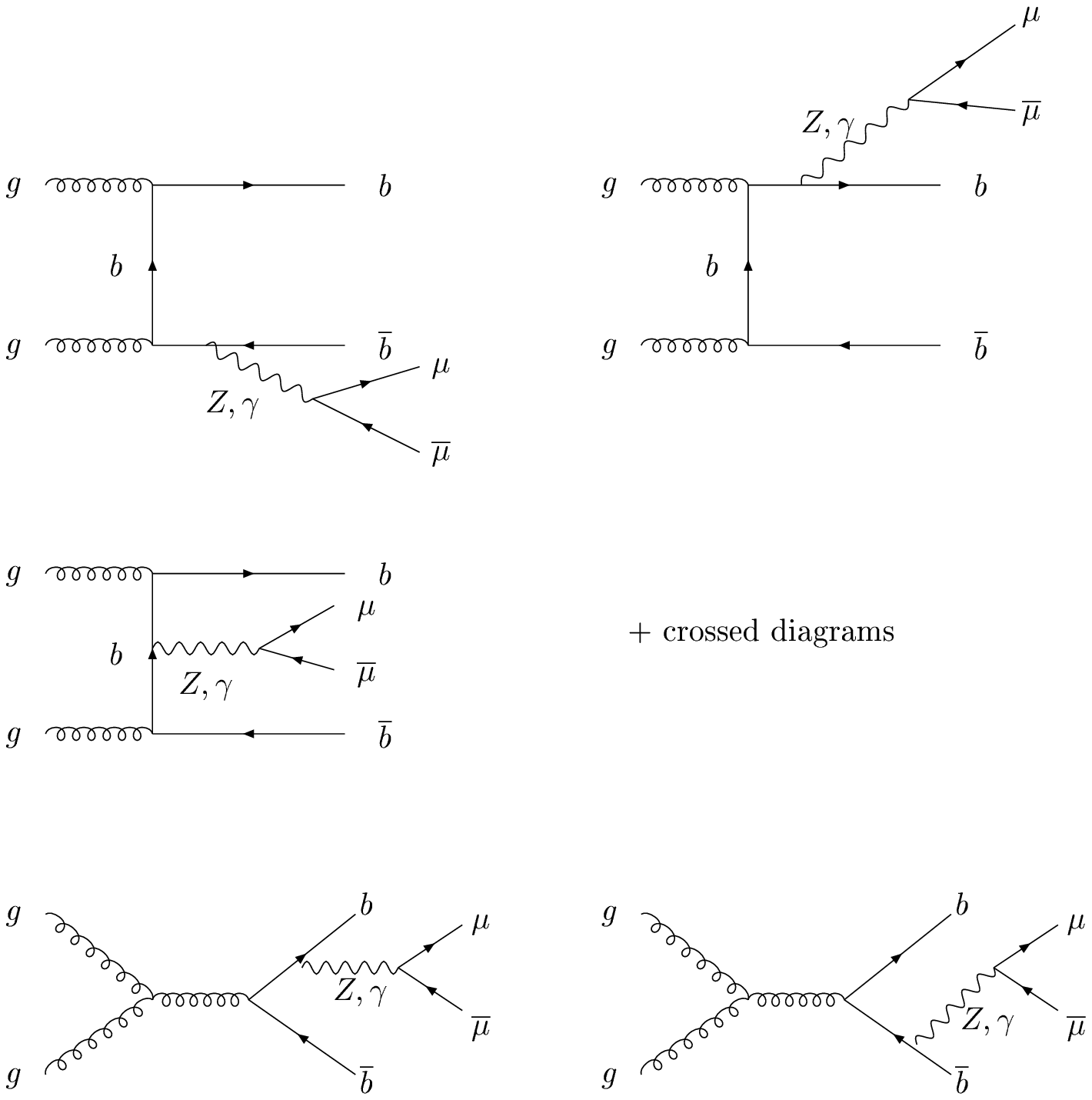}

\caption[]{
Feynman diagrams for $gg \to b\bar{b} \mu\bar{\mu}$.
\label{fig:Diagrams}
}\end{figure}
%


\begin{figure}
\centering\leavevmode
\epsfxsize=6in\epsffile{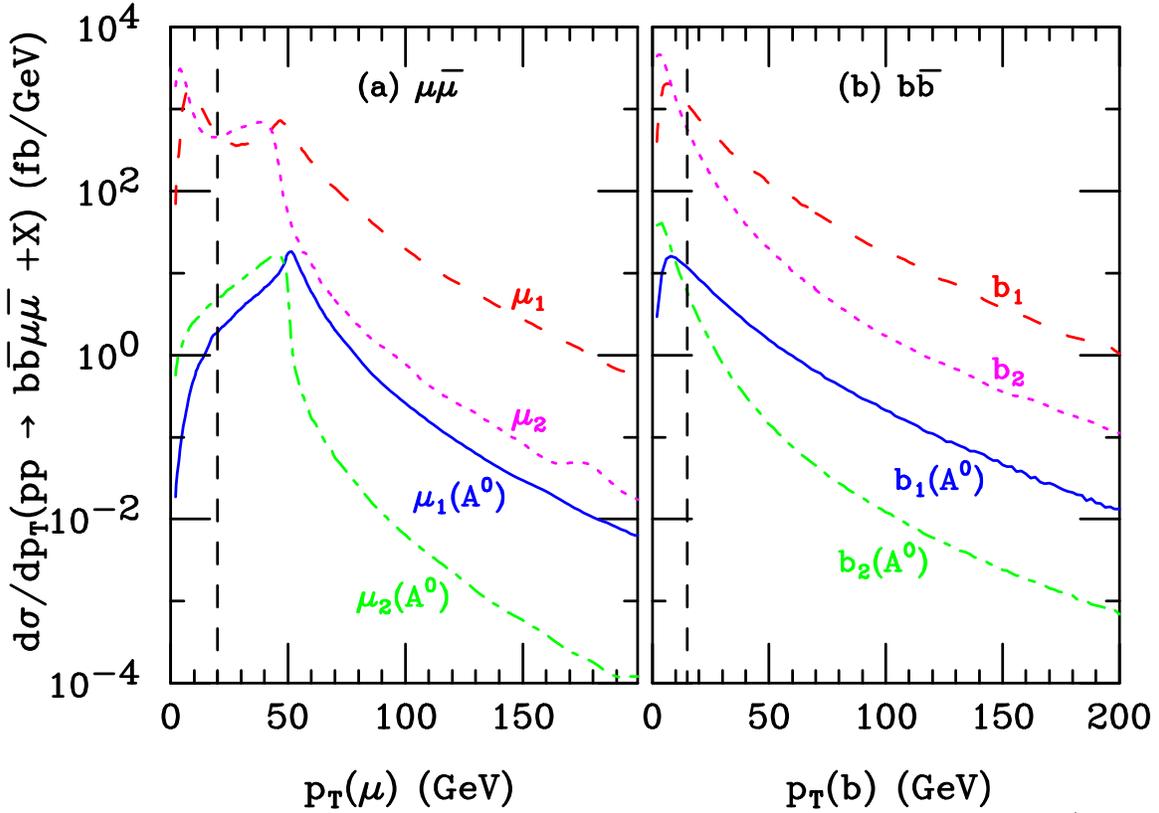}

\caption[]{
The $p_T$ distributions of muons and bottom quarks in fb/GeV 
for $pp \to b\bar{b} A^0 \to b\bar{b} \mu\bar{\mu} +X$ (solid and dot-dashed) 
with $m_A = 100$ GeV, $\tan\beta = 10$ and 
$m_{\tilde{q}} = m_{\tilde{g}} = \mu = 1$ TeV.
We have chosen $\mu_1$ or $b_1$ to be the muon or the b quark with higher 
transverse momentum.
Also shown is the contribution from the SM background of 
$pp \to b\bar{b} \mu\bar{\mu} +X$ (dashed and dotted) via
$gg \to b\bar{b} \mu\bar{\mu}$ and 
$q\bar{q} \to b\bar{b} \mu\bar{\mu}$.
The $p_T$ cuts remove events to the left of the vertical dashed line.
\label{fig:PTbm}
}\end{figure}


\begin{figure}
\centering\leavevmode
\epsfxsize=6in\epsffile{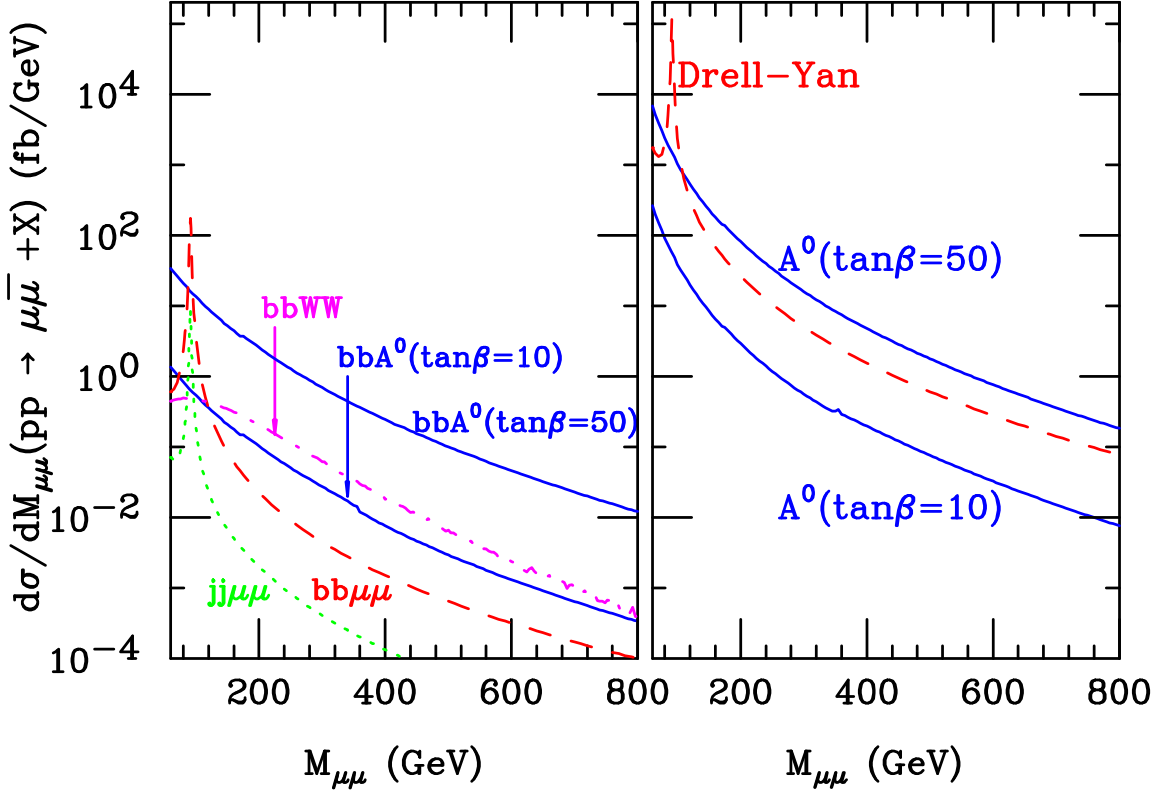}

\caption[]{

The invariant mass distribution ($d\sigma/dM_{\mu\bar{\mu}}$) for
$pp \to b\bar{b}A^0 +X \to b\bar{b}\mu\bar{\mu} +X$ 
at $\sqrt{s} = 14$ TeV, as a function of $M_{\mu\mu} = m_A$, for
$m_{\tilde{q}} = m_{\tilde{g}} = \mu = 1$ TeV and $\tan\beta = 10$ or 50.
Also shown are the SM backgrounds from
$pp \to b\bar{b}\mu\bar{\mu} +X$ (dashed),
$pp \to jj\mu\bar{\mu} +X$ (dotted), and
$pp \to b\bar{b}W^+W^- +X$ (dot-dashed).
The panel on the right shows the invariant mass distribution for the
inclusive final state $pp \to A^0 \to \mu\bar{\mu} +X$ and the Drell-Yan
background (dashed) with $M_{\mu\bar{\mu}} = m_A$.
We have applied the acceptance cuts as well as the efficiencies of b tagging
and mistagging for $L = 30$ fb$^{-1}$.
\label{fig:sigma}
}\end{figure}


\begin{figure}
\centering\leavevmode
\epsfxsize=6in\epsffile{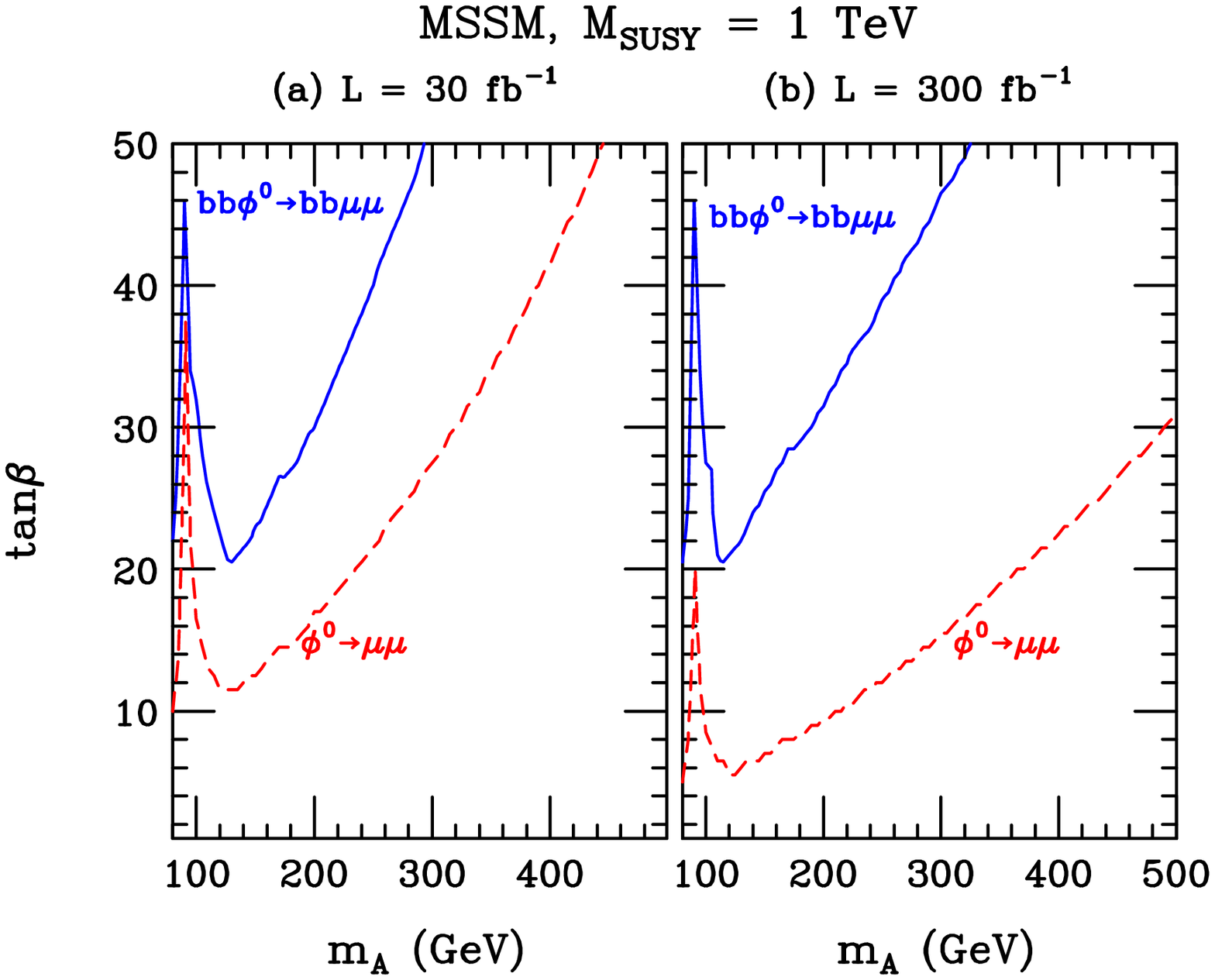}

\caption[]{
The $5\sigma$ contours at the LHC 
for an integrated luminosity ($L$) of 30 fb$^{-1}$ and 300 fb$^{-1}$ 
in the $m_A$ versus $\tan\beta$ plane.  
The signal includes $\phi^0 = A^0$ and $h^0$ for $m_A < 125$ GeV, 
but $\phi^0 = A^0$ and $H^0$ for $m_A \ge 125$ GeV.
The discovery region is the part of the parameter space above 
the $5\sigma$ contour.
\label{fig:contour}
}\end{figure}
%


\begin{references}
\bibitem{Guide} 
J.~Gunion, H.~Haber, G.~Kane and S.~Dawson, 
{\it The Higgs Hunter's Guide} (Addison-Wesley, Redwood City, CA, 1990).
\bibitem{MSSM}
H.P.~Nilles, Phys.~Rep. {\bf 110} (1984) 1; 
H.~Haber and G.~Kane, Phys.~Rep. {\bf 117} (1985) 75.
\bibitem{Model2}
J.~F.~Donoghue and L.~F.~Li,
Phys.\ Rev.\ D {\bf 19}, 945 (1979);
L.~J.~Hall and M.~B.~Wise,
Nucl.\ Phys.\ B {\bf 187}, 397 (1981).
\bibitem{Kunszt}
Z.~Kunszt and F.~Zwirner,
Nucl.\ Phys.\ B {\bf 385}, 3 (1992).
\bibitem{Richter-Was}
E.~Richter-Was, D.~Froidevaux, F.~Gianotti, L.~Poggioli, D.~Cavalli 
and S.~Resconi,
Int.\ J.\ Mod.\ Phys.\ A {\bf 13}, 1371 (1998).
\bibitem{Dandi}
T.~Garavaglia, W.~Kwong and D.-D.~Wu, Phys. Rev. {\bf D48}, 1899 (1993).
\bibitem{DGV}
J.~Dai, J.F.~Gunion and R.~Vega, Phys. Lett. {\bf B315} (1993) 355;
Phys. Lett. {\bf B345} (1995) 29; {\bf B387} (1996) 801. 
\bibitem{hbb}
E.~Richter-Was and D.~Froidevaux,
Z.\ Phys.\ C {\bf 76}, 665 (1997).
\bibitem{Nikita}
C.~Kao and N.~Stepanov, Phys. Rev. D {\bf 52} (1995) 5025.
\bibitem{CMS}
CMS Technical Proposal, CERN/LHCC 94-38 (1994).
\bibitem{ATLAS}
ATLAS Detector and Physics Performance Technical Design Report, 
CERN/LHCC 99-14/15 (1999).
\bibitem{mSUGRA}
A.H.~Chamseddine, R.~Arnowitt and P.~Nath,
Phys. Rev. Lett. {\bf 49} (1982) 970;
L.~Iba\~nez and G.~Ross, Phys. Lett. {\bf B110} (1982) 215;
R.~Barbieri, S.~Ferrara and C.~Savoy, Phys. Lett. {\bf B119} (1982) 343;
L.J.~Hall, J.~Lykken and S.~Weinberg, Phys. Rev. {\bf D27} (1983) 2359;
L.~Alvarez-Gaum\'e, J.~Polchinski and M.~Wise,
Nucl. Phys. {\bf B121} (1983) 495.
\bibitem{Vernon}
V.~D.~Barger and C.~Kao,
Phys.\ Lett.\ B {\bf 424} (1998) 69.
\bibitem{Tilman}
T.~Plehn and D.~Rainwater,
Phys.\ Lett.\ B {\bf 520}, 108 (2001).
\bibitem{Han}
T.~Han and B.~McElrath,
Phys.\ Lett.\ B {\bf 528}, 81 (2002).
\bibitem{Duane}
D.~Dicus and S.~Willenbrock, Phys. Rev. {\bf D39} (1989) 751;
D.~Dicus, T.~Stelzer, Z.~Sullivan and S.~Willenbrock,
Phys.\ Rev.\ D {\bf 59}, 094016 (1999);
C.~Balazs, H.~J.~He and C.~P.~Yuan,
Phys.\ Rev.\ D {\bf 60}, 114001 (1999).
\bibitem{CTEQ5}
H.~L.~Lai {\it et al.}  [CTEQ Collaboration],
Eur.\ Phys.\ J.\ C {\bf 12}, 375 (2000).
\bibitem{QCD1} 
%
S.~Dawson, Nucl. Phys. {\bf B359} (1991) 283; 
%
A.~Djouadi, M.~Spira and P.M.~Zerwas, 
Phys. Lett. {\bf B264} (1991) 440;
D.~Graudenz, M.~Spira and P.M.~Zerwas, 
Phys. Rev. Lett. {\bf 70} (1993) 1372.
\bibitem{QCD2} 
M.~Spira, A.~Djouadi, D.~Graudenz and P.M.~Zerwas, 
Nucl. Phys. {\bf B453} (1995) 17. 
\bibitem{QCD3} 
S.~Dawson, A.~Djouadi and M.~Spira, 
Phys. Rev. Lett. {\bf 77} (1996) 16.
\bibitem{QCD4} 
S.~Catani, D.~de Florian and M.~Grazzini,
JHEP {\bf 0201}, 015 (2002);
R.~V.~Harlander and W.~B.~Kilgore,
Phys.\ Rev.\ Lett.\  {\bf 88}, 201801 (2002);
C.~Anastasiou and K.~Melnikov,
arXiv:hep-ph/0207004.
\bibitem{Plumper} 
B.~Plumper,
DESY-THESIS-2002-005.
\bibitem{Manuel} 
E.~Braaten, J.P.~Leveille, Phys. Rev. {\bf D22} (1980) 715; 
M.~Drees and K.~Hikasa, Phys. Lett. {\bf B240} (1990) 455; 
(E)-$ibid.$ {\bf B262} (1991) 497.
\bibitem{Madgraph}
MADGRAPH, by T.~Stelzer and W.F.~Long,
Comput. Phys. Commun. {\bf 81}, 357 (1994).
\bibitem{Helas}
HELAS, by H.~Murayama, I.~Watanabe and K.~Hagiwara,
KEK report KEK-91-11 (1992).
\bibitem{HGG} 
H.~Baer, M.~Bisset, C.~Kao and X.~Tata,
Phys.\ Rev.\ D {\bf 46}, 1067 (1992).
\bibitem{Brown} N.~Brown, Z. Phys. {\bf C49} (1991) 657.
\bibitem{HZ2Z2}
H.~Baer, M.~Bisset, D.~Dicus, C.~Kao and X.~Tata,
Phys.\ Rev.\ D {\bf 47} (1993) 1062;
H.~Baer, M.~Bisset, C.~Kao and X.~Tata,
Phys.\ Rev.\ D {\bf 50} (1994) 316.
%
\end{references}
\end{document}